\documentclass[twoside]{article}
\usepackage{fleqn,espcrc2,epsfig}

\newcommand{\slB}{\raise.15ex\hbox{$/$}\kern-.57em\hbox{$B$}}
\newcommand{\slC}{\raise.15ex\hbox{$/$}\kern-.57em\hbox{$C$}}
\newcommand{\slCP}{C\raise.15ex\hbox{$/$}\kern-.57em\hbox{$P$}}
\newcommand{\be}{\begin{equation}}
\newcommand{\ee}{\end{equation}}
\newcommand{\bear}{\begin{eqnarray}}
\newcommand{\ear}{\end{eqnarray}}

\title{Electroweak Baryogenesis in Supersymmetric
Variants\thanks{Heidelberg Preprint HD-THEP-01-4; presented
at the SUSY 30 workshop, October 2000, Minneapolis}
}
\author{Michael G. Schmidt\\
Institut f\"ur Theoretische Physik\\ Universit\"at Heidelberg\\
Philosophenweg 16, D-69120 Heidelberg, Germany}

\begin{document}
\begin{abstract}
We argue that the creation of a baryon asymmetry in the early universe
is an intriguing case where several aspects of
``Beyond'' physics are needed. We then concentrate on baryogenesis
in a strong first-order phase transition and discuss  that supersymmetric
variants of the electroweak theory (MSSM and some version of NMSSM)
rather naturally provide the necessary ingredients. The
charginos and the stops play a prominent role. We present
CP-violating dispersion relations in the chargino sector and show
results of a concrete model calculation for the asymmetry production
based on quasi-classical Boltzmann transport equations and sphaleron
transitions in the hot electroweak phase.\vspace{1pc}
\end{abstract}
\maketitle
\section{Introduction}

The great progress in elementary particle physics in the last decades
was based on high energy accelerator experiments and on the detection
that all the newly found particles and their interactions can be described
by a rather simple set of gauge theories -- the Standard Model (SM) of
elementary particle physics.

The Big Bang theory of cosmology predicts very high temperatures
in the early universe. The elementary particles -- their spectrum and
interactions -- become essential, and such processes in the hot plasma can be
calculated in the SM. This led to a detailed picture (``the first three
minutes'') one could only dream of not so long ago. Still there are some
fundamental problems of standard cosmology related to initial conditions:
starting the history of the universe with an epoche of exponential
growth today is the only convincing solution -- in particular since it
also generates the right type of fluctuations seen in
the background radiation and needed for galaxy formation. There is also the
finding that a large amount of dark matter not present in the SM
is needed to stabilize gravitationally the observed structures and
to explain the timescale of galaxy formation. This requires physics
beyond the SM. In spite of the success of the SM there are also good
reasons inside elementary particle physics to require extensions
of the SM. We mention the problem of scales and the Grand Unification
including gravity. Such ``Beyond the SM'' models may be very severely
restricted by the requirement of a reasonable cosmology. The explosion
in astrophysical observations based on telescopes, satellites, and
fully computerized evaluation of a huge material made this point more
and more relevant.

Up to now there is no direct evidence for ``Beyond the SM'' physics
and thus we have to cultivate theoretical prejudices, building models,
and to ask for observable consequences in accelerator experiments
but also in astrophysics. Supersymmetry \cite{1'}
is the most attractive theoretical
idea in the last decades which has the chance to be tested in forthcoming
experiments. It is de facto also a very natural ingredient of string
theory -- the modern version of a grand unification including  gravity.
Supersymmetry is theoretically appealing because it just slightly extends our  
ideas about geometry. It is also attractive phenomenologically: the chiral
superfields combine quarks and leptons with interesting scalar fields,
in particular the strongly coupled stops, and lead to a rich  
Higgs/Higgsino/gaugino spectrum. There are convincing dark matter candidates.
It produces naturally potentials with flat directions and -- not to forget
the oldest pro-argument -- the stabilization of a chosen fine-tuned
scale. A light Higgs (114 GeV?) would be strongly in favor of supersymmetry.
Unfortunately the choice of supersymmetric models is not unique and
the present discussion -- different from the early days of SUSY \cite{23} --
is mostly on the minimal (MSSM) version because even there the new
parameters related to SUSY breaking are hard to fix.

Baryogenesis, the generation of an asymmetry between baryons and antibaryons
in the early universe, is a very important chapter of cosmology. Even though
this is a tiny effect $(n_B/n_\gamma\sim 10^{-10})$ the ``tiny''
rest of baryons left after pair-annihilation constitutes our world!
Following Sakharov \cite{1}, it requires a few highly nontrivial effects to
combine to produce such an asymmetry. (i) Baryon number B violation
is very natural in grand unified models, but it also happens in the
electroweak gauge theory via instantons or in high temperature physics
via sphaleron thermal transitions where, however, (B-L) number
is conserved. At very high temperatures  the Higgs field does
not have a quasi-classical vev, and this B+L-violating  transition is
unsuppressed.  It can be estimated very accurately \cite{2}. (ii) CP has to
be violated. In the SM we have CP-violation in the KM-matrix, but this is
a very small effect. In ``Beyond'' models like the MSSM there
are more possiblities, but the size of such effects is still under
discussion \cite{24,11}. (iii) One needs nonequilibrium. In the very early  
universe at the
(GUT/inflation) scale the expansion of the universe provides strong
nonequilibrium, but later on at the electroweak scale
$(T\sim$ 100 GeV) one presumably needs a strong first-order phase
transition (PT)  which we know e. g. from the condensation of vapor.

Such a PT between a symmetric phase and a ``Higgs''
phase, where the Higgs field
gets its quasiclassical vev, is naively predicted at the electroweak
scale in the early universe if one uses thermal perturbation theory
with a positive Higgs thermal (mass)$^2$ (``hard thermal loops'').
Thus the (Sakharov) necessary conditions for baryogenesis could
be fulfilled in the SM \cite{3}. If a GUT interaction conserves B-L,
no B-L would be
created during pre/reheating after inflation, and B+L would be washed
out in the (quasi) equilibrium period before the electroweak PT. The
last chance to create an asymmetry would then be during this PT. However,
in the last few years it became clear \cite{4}
that there is no PT at all for
the SM with a Higgs mass $m_H$ larger than the $W$-mass and thus
in the experimentally not excluded range above $\sim$ 110 GeV.

One then has to return to other sources of a baryon asymmetry. In
particular ``leptogenesis'' \cite{5} in SO(10) models with a B-L-violating
interaction is rather popular because it allows for some
(loose) connection to neutrino-mass generation. But one can also
discuss other directions, e.g. Affleck-Dine condensate
instability and Q-Balls \cite{6}
(squark droplets), or electrogenesis \cite{7}
with an out-of equilibrium right-handed electron. Staying with the
electroweak PT one has to discuss non-standard variants -- and this we
will do in the following. Quite generally we can conclude that any attempt
to explain the baryon asymmetry requires non-standard model ingredients.
Thus baryogenesis is a very interesting laboratory for such models.

\begin{figure}[t] 
\begin{picture}(100,110)
\put(40,0){\epsfxsize4cm \epsffile{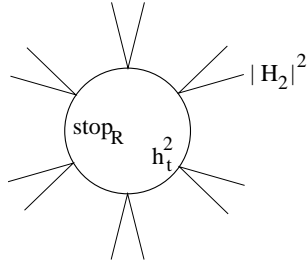}}
\end{picture} 
\caption{$(H_2^*H_2)^{3/2}$-contribution of the stop in one loop order.}
\label{fig_1}
\end{figure}

\section{Supersymmetric Modifications of the SM and the Electroweak
Phase Transition}

In order to increase the strength of the electroweak PT, one has to
provide a larger ``$\varphi^3$'' term in the
effective high temperature
Higgs action (three-dimensional):

(I) In the MSSM a ``light'' stop$_R$, superpartner of the r.h. top, with a
big Yukawa coupling to a Higgs gives such a $-(\varphi^*\varphi)^{3/2}$  
potential (fig. 1). ``Light'' means that its (mass)$^2$ in the symmetric
phase  given by
\be\label{1}
m^2_3=m_{SB}^2+m^2_{T}\ee
where $m^2_{SB}$ is the SUSY breaking scalar (mass)$^2$ and
$m^2_{T}\sim (gT)^2$ the thermal mass, is small -- this is
for negative $m^2_{SB}$! At zero temperature one has $m^2_{stop_R}=
m^2_{SB}+m^2_{top}$, i.e. its mass is below that of the top. One can discuss
the Higgs/stop system perturbatively \cite{8,9} and on the lattice \cite{10}
starting from a 3-dimensional effective action obtained by integrating out
the massive and nonstatic degrees of freedom. In both approaches one
finds (for a strong PT nonperturbative effects near $\varphi\sim0$
turn out to be not so important!) that there is a strong first-order PT
even at lightest SUSY Higgs masses as high as 110 GeV
and for $stop_R$-masses
in the range 160 GeV $\leq m_{stop_R}\leq m_{top}$. This depends
on the $stop_L$-mass, on $\tan\beta$, and on $m_{A_0}$, the Higgs
bound might be a few GeV higher for large $\tan\beta$. A detailed
analysis can be found in ref. \cite{11,12}. The originally found
two-stage PT \cite{8} at even smaller $m_{stop_R}$ including a color-breaking
phase seems to be excluded because in this case one
never would return to the Higgs phase \cite{13}.

\begin{figure}[t] 
\begin{picture}(100,110)
\put(-70,-345){\epsfxsize14cm \epsffile{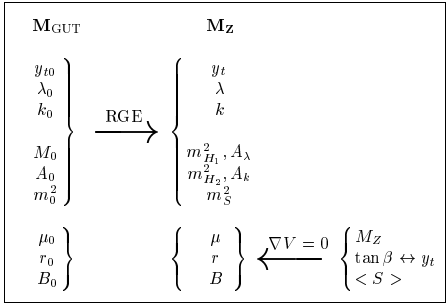}}
\end{picture} 
\caption{Sketch of our procedure to determine the weak scale
parameters from the GUT parameters.}
\label{fig_2}
\end{figure}

(II) In supersymmetric models including a gauge singlet superfield
(``next to minimal'', NMSSM) a ``$\varphi^3$''-type term
is already present on the tree level. We consider a superpotential
\cite{16,14,15}
\be\label{2}
W=\lambda SH_1H_2+\frac{k}{3}S^3+\mu H_1H_2+rS\ee
with soft SUSY breaking scalar masses, gaugino masses and with A-terms
\bear\label{3}
&&{\cal L}_A=\lambda A_\lambda SH_1H_2+\frac{k}{3}A_kS^3
\\
&&+\ {\rm
usual\ (squark,\ slepton,\ Higgs)\ A-terms}\nonumber\ear
The first term in (\ref{3}) or the term
$\lambda^*\mu S^*H_1H_1^*$ from (\ref{2}) e.g. are such ``$\varphi^3$''-terms
if the potential leads to vevs of the singlet $<S>$ comparable
in size with $<H_{1,2}>$, the electroweak scale.

\begin{figure}[t] 
\begin{picture}(100,110)
\put(-90,-275){\epsfxsize12cm \epsffile{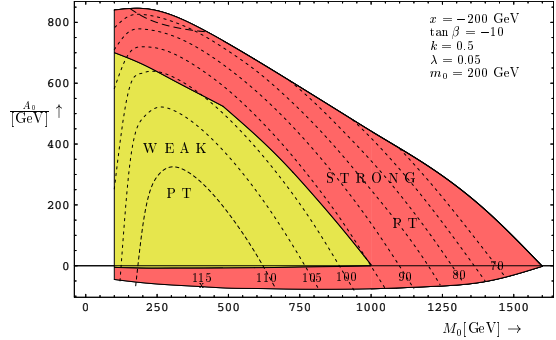}}
\end{picture} 
\caption{Scan of the $M_0$-$A_0$ plane for a set of $(x=<\!\!\!S\!\!\!>,\tan\beta,k)$:
strongly first order means $v_c/T_c>1$. Dotted lines
are curves of constant mass of the lightest CP-even Higgs boson. In the region
above the dashed line the lightest Higgs is predominantly a singlet.}
\label{fig_3}
\end{figure}

The superpotential (\ref{2}) is not $Z_3$-symmetric any more like its
first two terms. Thus there are no problems with domain
walls \cite{15'} in case
this $Z_3$ is spontaneously broken. We also can find easily parameter
sets where the particle spectrum is not in contradiction with
experiments and the size of the singlet $S$ field is of the
order of the electroweak scale in the relevant region of the potential
in $S,H_1,H_2$. In the original version \cite{17} one always obtains
$<S>\gg <H_1,H_2>$. However, we had to introduce a $\mu$-term again and
thus we have finetuning problems as in the MSSM \cite{18}. (Indeed to avoid
these was one of the reasons to introduce a singlet!). Besides
this there is the danger of quadratically exploding singlet
tadpoles \cite{19,20}.
Such tadpole diagrams require three ingredients: (i) a singlet
field, (ii) renormalizable interactions, (iii) soft SUSY breaking
terms. One can try to forbid the dangerous nonrenormalizable operators,
e.g. in models \cite{20,21}
with gauged $R$-symmetry  or duality symmetry, both
broken at some superheavy scale. This of course still leaves
us with the $\mu$-finetuning problem. Another way to evade the
tadpole divergencies restricts the SUSY breaking terms thus avoiding
destabilization via the tadpole: Gauge-mediated SUSY breaking
(GMSB) in the context of singlet models does not have domain wall
problems. A $\mu$-parameter is generated by radiative corrections
and the singlet vev \cite{22}. However, one of the properties of
GMSB models seems to be the strong suppression of A-terms, important
for the creation of a ``$\varphi^3$''-term as we just argued before.

The model (\ref{2}), (\ref{3}) different from elegant models in the old
days \cite{23}
contains quite a few parameters
even if one introduces universal SUSY breaking scalar $m^2_0$
and gaugino masses $M_0$ at the GUT scale. One can connect the
parameters at the GUT and the electroweak scale via a set of
renormalization-group equations. Indeed we found \cite{14,15}
a way to select
cases where $M_z, \tan\beta$ and $<S>$ are fixed from the outset.
The remaining parameters are $\lambda, k$ in (2) and universal $m^2_0,M_0,
A_0$ (fig. 2). The parameter set is restricted by the postulate
of a stable electroweak minimum and by an experimental restriction
to chargino masses $\geq$ 90 GeV. We then found a strong first-order
PT $(v(T_{crit.})/T_{crit.}\geq1)$ for
light CP even Higgs masses as large as 115 GeV (and even higher)
(fig. 3).

In both models I and II CP-violation is much less restricted than in the SM.  
In the MSSM one can have explicit CP-violating
phases in the $\mu$ and $A_{top}$ parameter
\cite{24,11}, in the NMSSM
there are even more possibilities. They are restricted  by the
experimental bound on the neutron electric dipole moment.
In the NMSSM \cite{25} --
not in the MSSM \cite{25,26}-- we found also the very appealing possibility
of a spontaneous CP violation in the $H_1,H_2,S$ system
considering the effective field equations at the high temperature
of the PT in the phase transition region (``bubble wall''). This
CP violation can be strong without being limited by experimental
bounds.

If there is a first-order PT, one can discuss analytically
and numerically the shape of the critical bubble, calculate
the transition probability and the nucleation temperature
(``one bubble/universe''). The Higgs phase bubble expands, and
due to the friction of the hot plasma, it approaches a
stationary expansion with a ``wall''-velocity $v_w$. This is the
most interesting period where the baryon asymmetry is supposed to be
generated: There is a strong baryon number violation due to
the ``hot'' sphaleron transition in the symmetric phase in front
of the bubble wall; there is CP violation in the bubble wall
region and there is strong equilibrium due to the  wall
sweeping through the hot plasma and changing the masses of many particles
since it is a Higgs field configuration. Of course these are just
necessary conditions which are well fulfilled: They have to be
bundled into a concrete scenario of baryon asymmetry formation.

\begin{figure}[t] 
\begin{picture}(100,110)
\put(5,0){\epsfxsize7cm \epsffile{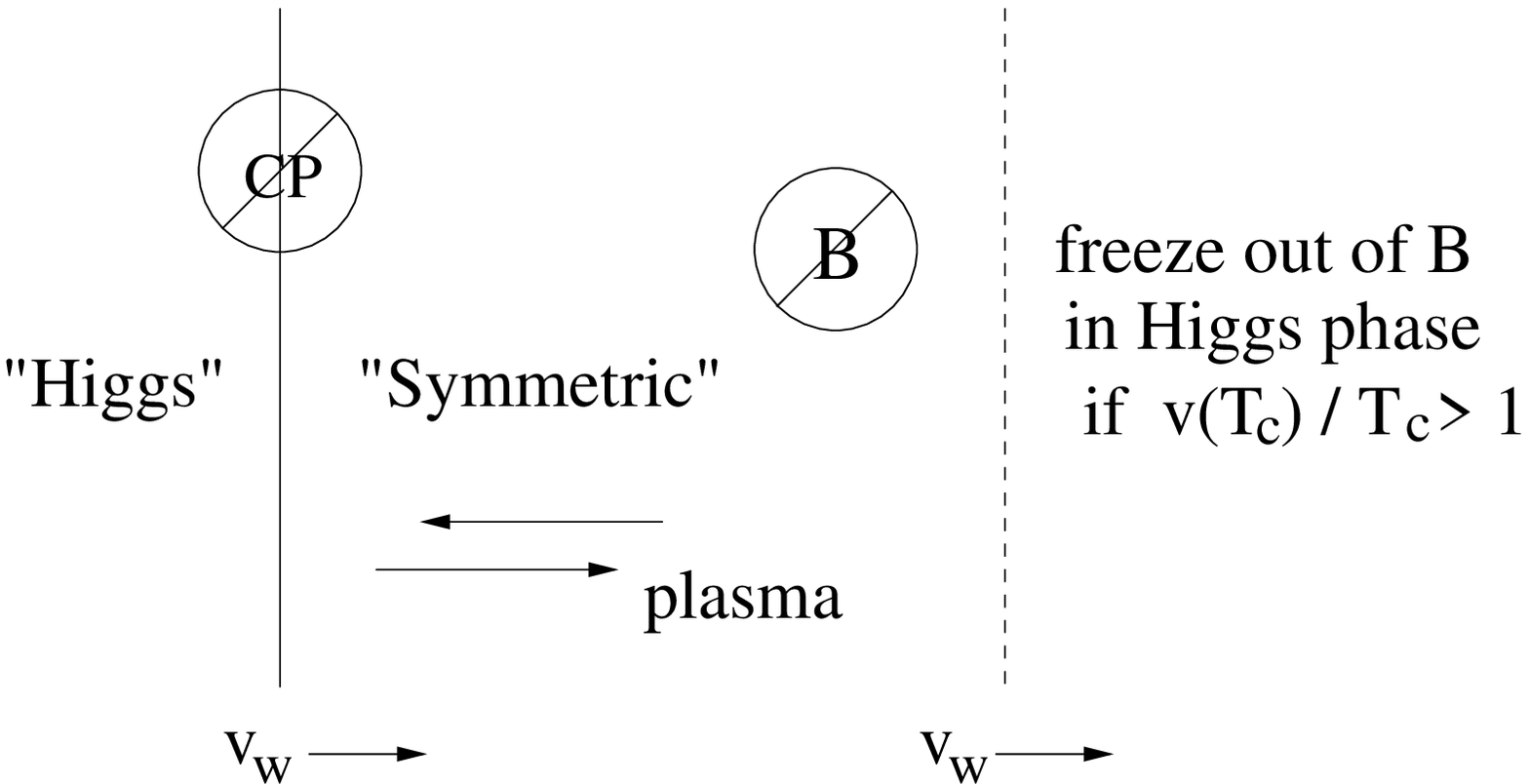}}
\end{picture} 
\caption{}
\label{fig_4}
\end{figure}

\section{Baryogenesis}

We consider the following model \cite{15,26,27,28,29,29'}
for the generation of a baryon asymmetry
(fig. 4): The bubble wall of the first-order PT proceeds with
stationary velocity $v_w$; the particles in the hot plasma which is in
thermal equilibrium interact with the Higgs fields $H_{1,2}(z)$ of the wall,
where $z$ is the direction perpendicular to the wall, i.e. they
change mass, and thus nonequilibrium is created. In case of CP violation,
particles and their CP conjugate antiparticles have different
dispersion relations. This creates an asymmetry -- still not the
baryon asymmetry-- which is transported in the region in front of
the bubble wall where the interactions in the hot plasma transform this
asymmetry e.g. of charginos and anticharginos into an asymmetry
between left-handed quarks and their antiparticles. The latter
than creates a baryon asymmetry through thermal sphaleron transitions
in the hot symmetric phase. These transitions are not in equilibrium,
otherwise the asymmetry would be reduced again. After some time the
Higgs-phase behind the bubble wall takes over. The usual Higgs-phase
sphaleron should be ineffective now in order to freeze out the
baryon asymmetry just generated. This is why we need a strong
first-order PT with the Higgs vev $v(T_c)\stackrel{\scriptstyle
>}{\sim} T_c$ at the critical temperature: The sphaleron transition
is suppressed now by a factor $\sim \exp(-v(T_c)/T_c)$.

For ``thick'' bubble wall profiles with a typical length scale
$L_W\gg 1/T$ -- and this happens to be the case in our models --
most of the particles have thermal momenta $p\sim T\gg 1/L_W$
and behave semiclassically. They can be treated using the WKB
approximation \cite{15,27,28}. There is a small thermal relaxation
time $\ll L_W$. It also turns out to be a reasonable
approximation to neglect thermal nonequilibrium
even in the wall region and just to discuss diffusion equations
for the chemical potentials of particles minus antiparticles.
In the case of our supersymmetric models the chargino system with mass
matrix
\be\label{4}
(i\tilde W^-,\ \tilde h_1^-) {\ \ \ M_2 \qquad \ g_2(H^0_2)^*
\choose g_2(H^0_1)^*\ \mu+\lambda S}{i\tilde W^+ \choose \tilde h_2^+}
\ee
is most important. After mixing the lightest charginos with mass
$m(z) e^{i\theta (z)}$and their CP conjugates we have up to order
$\hbar$ dispersion relations \cite{15,28}
\bear\label{5}
&&E=(\vec p^2+m^2(z))^{1/2}\\
&&\pm\frac{1}{2}(\theta'(z)+\delta'(z)\sin^2 b-\gamma'\sin^2
a)m^2/(\vec p^2+m^2)\nonumber\ear
Here the prime mass is the derivative perpendicular to the wall,
$\delta$ and $\gamma$ are phases appearing in the diagonalization
matrices, $\theta$ is the phase of the diagonalized Dirac mass
and $\sin^2a, \sin^2b$ contain the parameters of the chargino
system (\ref{4}). $\vec p$ in (\ref{5}) is the kinetic momentum
related to the group velocity $\frac{\partial \omega}{\partial
\vec p_{can}}$. It differs from the canonical momentum $\vec p_{can}$
because of the CP-violating terms. This is the physical momentum
which should enter quasi-classical transport equations \cite{28}.
(\ref{5})
is symmetric under the exchange of $H_1$ and $H_2$. Because of the derivatives
$\theta', \gamma', \delta'$ CP violation only becomes relevant in the
wall region. The particles with such quasiclassical dispersion relations
-- here we mention the charginos and their
superpartners and the particles they mainly interact with (stop/top) --
are now used in classical Boltzmann transport equations. The light
quarks only enter via a strong $(SU(3)_c)$ sphaleron interaction.
The derivative terms in (\ref{5}) then constitute the only source terms
in the Boltzmann equations for the difference between particles and
their CP conjugates needed for the asymmetry.
The time derivatives and $z$-derivatives in the
Boltzmann equations can be substituted by a
$\bar z=z-v_wt$ derivative in the
stationary case.

The equations are
solved in the fluid approximation (for bosons/fermions)
\be\label{6}
f_i(\vec x,\vec p,t)=1/\{e^{\beta(E_i-v_ip_z-\mu_i)}\pm1\}\ee
where as mentioned before we have included as small perturbations
from the equilibrium distribution only the chemical potentials $\mu_i$ and
(for consistency) flux velocities in the plasma. We expand in these small
quantities (linear response) as well as in the wall velocity
to first order. Taking momenta of the transport equations
and eliminating $v_i$, one arrives at a set of diffusion equations
which can be solved partly analytically, partly numerically.

If one writes Boltzmann equations just for particles \cite{30,31}, the
main source term is related to the $z$-dependent mass due to the wall,
and CP-violating effects are not important. A balance between the
pressure of the bubble and the friction due to the plasma leads
to a self-consistent wall velocity $v_w$. Indeed,
in our recent evaluation it turns out
that the stop plays an important role in reducing the
wall velocity compared to the SM (fig. 5) to values of the
order 0.01. This helps technically in the expansion,
but more importantly it  turns out to strengthen the final baryon
asymmetry, as we will see now.

\begin{figure}[t] 
\begin{picture}(100,140)
\put(0,0){\epsfxsize7cm \epsffile{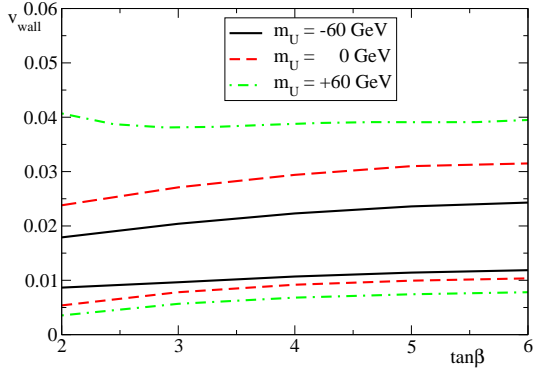}}
\end{picture} 
\caption{Wall velocity in dependence on the parameter $\tan\!\beta(T\!=\!0)$
for $m_Q=2$TeV, $A_t=\mu=0$, and $m_A=400$GeV for $m_U^2=m_{SB}^2=-60^2, 0,
60^2$GeV${}^2$. Lower bunch of graphs for $\delta'=0$, upper for
$\delta'\neq 0$ (exact linear response), see ref. \cite{31}.}
\label{fig_5}
\end{figure}

The final baryon asymmetry is obtained as \cite{28,15}
\be\label{7}
\eta_B=n_{B/S}=\frac{135 \Gamma_{WS}}{2\pi^2 g_*v_WT}\int^\infty_0
d\bar z \mu_{B_L}(\bar z) e^{-\nu\bar z}\ee
with $\nu=\frac{3\cdot48}{7} \Gamma_{WS}/2v_W$, and where $\Gamma_{WS}$ is the
electroweak sphaleron rate in the hot phase, $g^*$ the
effective number of degrees of freedom; the integration is over
the z-region in front of the  wall and $\mu_{BL}$
is the chemical potential for left-handed baryonic matter
minus its CP cojugate.

In fig. (6)  we present a typical result \cite{29'} for the baryon asymmetry
in the MSSM due to the symmetric combination of Higgsino sources
-- only this one is nonvanishing using the dispersion relations
for quasiclassical particles with kinetic momentum. (The antisymmetric
combinations proportional to $(\tan \beta)'$ as well as stop sources
vanish in this approach. They are, however, present in the diffusion
equations obtained in the spirit of quantum-Boltzmann equations
\cite{29,12} or if one uses canonical momenta in the WKB approach
\cite{15}.) Thus one needs rather strong CP violation -- fig. (6) is
for the case of maximal CP violation -- in order to get close
to the observed baryon asymmetry. In the NMSSM such a strong CP
violation naturally appears in the singlet sector due to
``transitional'' spontaneous CP violation.

\section{Summary}

Baryogenesis is a very challenging problem at the borderline
between cosmology and particle physics because of its highly
nontrivial necessary ingredient. Independent of concrete models
its explanation will always involve physics beyond the SM! We have
discussed sypersymmetric models leading to a strong first-order PT.
For this the MSSM requires a Higgs mass below 110 GeV-115 GeV (depending
on its parameters) and thus is at the borderline to be ruled out
by experiments, the NMSSM in the version presented here is more
flexible, but again it is testable in accelerator experiments;
like in all SUSY models one needs a rather lowlying Higgs.
We consider this to be an advantage compared to models in the far
away GUT region.

\begin{figure}[t] 
\begin{picture}(100,140)
\put(20,-70){\epsfxsize6.5cm \epsffile{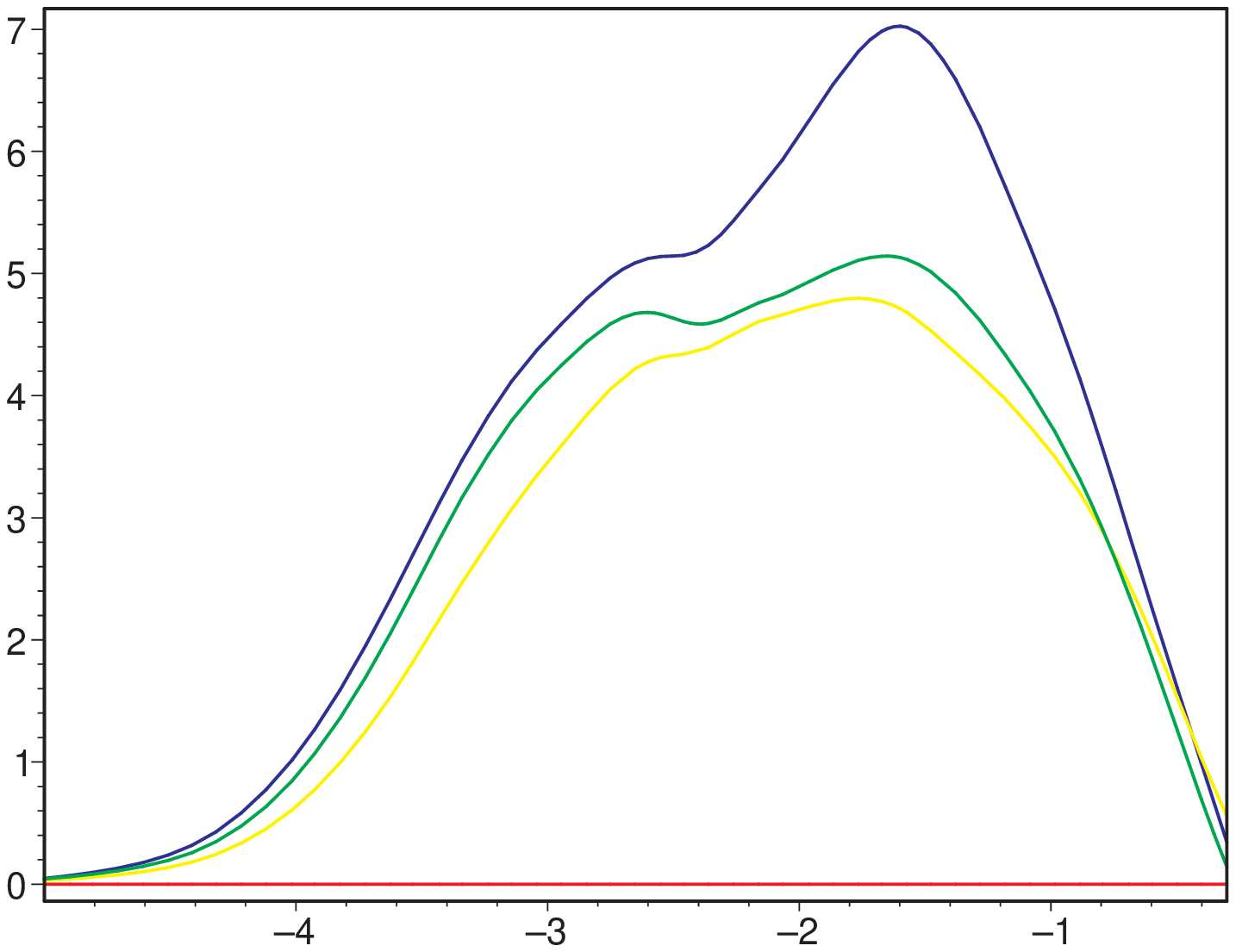}}
\put(110,-20){$\ln_{10}(v_w)\rightarrow $}
\put(-370,165){$\frac{\eta_B}{2\times10^{10}}\uparrow$  }
\put(45,115){{$H_1+H_2$}}
\put(0,85){{$\eta \uparrow$}}
\end{picture} 
\caption{$H_1+ H_2$  contributions to the baryon asymmetry dependent
on $v_w$ for different values of the wall thickness $L_w=20/T,15/T,10/T$
(from below) and $|\mu|=|M_2|=150$ GeV,
arg$(\mu M_2)=\pi/2$ and $\delta\beta=0.01$. $\eta$ is given in units
of $2\times 10^{-11}$ (observational bound).}
\label{fig_6}
\end{figure}

\section*{Acknowledgement}
I would like to thank S. Huber and P. John for
enjoyable collaboration on the subjects of this talk and N. Polonski
and S. Weinstock for discussions.

\end{document}